# Augmenting Customer Journey Maps with quantitative empirical data: a case on EEG and eye tracking

**Rui Alves, Veranika Lim, Evangelos Niforatos**
Madeira-ITI, University of Madeira
Polo Científico e Tecnológico da Madeira,
Caminho da Penteada, 9020-105 Funchal
{rui.alves|veranika.lim|evangelos.niforatos}
@m-iti.org

**Monchu Chen, Evangelos Karapanos, Nuno Jardim Nunes**
Madeira-ITI, University of Madeira
Polo Científico e Tecnológico da Madeira,
Caminho da Penteada, 9020-105 Funchal
{ekarapanos|monchu|njn}@uma.pt

## ABSTRACT
This paper introduces the use of electroencephalography (EEG) and eye tracking in exploring customer experiences in service design. These tools are expected to allow designers to generate customer journeys from empirical data leading to new visualization methods and therefore improvements in service design deliverables.

### Author Keywords
Eye tracking, Electroencephalography, Customer Journey, Touchpoint, Service Design.

### ACM Classification Keywords
H5.m. Information interfaces and presentation (e.g., HCI): Miscellaneous.

## INTRODUCTION
The increasing emphasis on creating meaningful and memorable customer experiences [3], has led service design to become an important application area within Human-Computer Interaction. One of the most popular and widely accepted design tools for service design is the customer journey, a sequential visualization of all possible touchpoints of a service with customers [6]. According to Holmlid and Evenson [4], a customer journey is a "walk in the customer's shoes", in an attempt to depict customer experiences while going through a service. A generally accepted definition of touchpoint is a point or moment of contact between an organization or brand and a customer or stakeholder [8]. Touchpoints consist of a cluster of artifacts (tangible) and activities (human interactions) that happen in a specific geographical or physical space (environment) involving a frontstage, a backstage and several lines of visibility. The customer journey map typically represents only the frontstage from the users' perspective, while a service blueprint covers detailed activities in the backstage which can be modeled by business and systems oriented techniques, such as Business Process Modeling Notation (BPMN). The backstage provides information and/or resources required by the frontstage, while the activities in the frontstage are visible and accessible to the customer and therefore critical to form the customer experience. A limitation of the frontstage is the lack of visualization and formal description of user experience, i.e. modeling, tools. Overall, current customer journey visualizations deeply rely on non-empirical and qualitative data, thus risking being subjective.

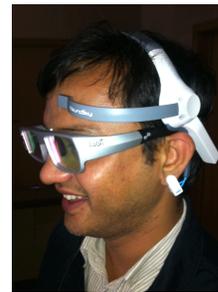

**Figure 1. A user wearing the mobile eye tracker and the wearable EEG.**

### Augmenting customer journeys with empirical data
While customer journeys have been a widely used tool in service design, they lack grounding on empirical data. Our approach augments customer journeys with relevant quantitative metrics of users' experiences that are captured in field studies.

In an initial case study with four participants, we explored the use of two bio-sensing devices: a) mobile eye tracker and b) wearable EEG. Attention levels, as gathered from EEG, and eye behavior seem to reveal a certain profile, exhibiting implicit human awareness during the experience of a service [10]. We have used Neurosky MindWave [6], an EEG wearable device, to measure attention levels, as well as neural oscillations such as alpha waves. A Tobii glasses mobile eye tracker [9] was used to collect eye gaze, fixations, and eye blink frequency. These measurements indicate customers' interests in particular artifacts and provide insights into their cognitive and affective state, thus offering service designers richer information to ground their design decisions. The duration and frequency of eye blinks, for example, have been proven to inform about workload levels [1, 2]. Therefore, these tools are expected to support service innovation and improvements leading service designers to more meaningful and tangible business decision-making. Collected eye blinks and attention levels are annotated in the customer journey visualizations. Richer data (such as the full video stream with gaze behavior) may further support reasoning about particular moments.

*Visualizations*

Visualizations help service designers to formulate insights from the user material collected, to communicate these insights to their clients and as a way of keeping the data alive [7]. An augmented customer journey could inspire better communication across designers and ground design decisions based on quantitative data. On the other hand, raw data such as the exact gaze behavior of an individual or the video stream of his visual field may provide rich insights in the perceptual and cognitive processing of surround stimuli. Shifting across either these levels or empirical data representation is crucial to design activity [5].

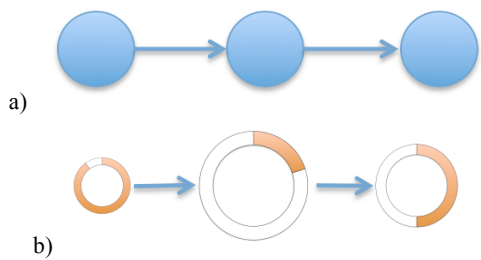

a)

b)

**Figure 2. Customer Journey visualizations:
a) Standard representation of each touchpoint;
b) Circle radius indicating attention level,
in the outer ring the normalized eye blinks.**

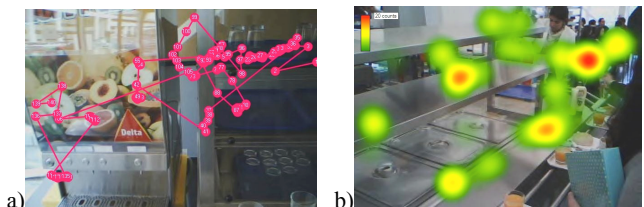

a)     b)

**Figure 3. Eye gaze behavior:
a) Fixation representation; b) Heat map.**

## DISCUSSION

This paper introduces a concept of visualizing quantitative data through the customer journey. In contrary to other approaches, we propose the annotation of EEG and eye tracking data on traditional customer journey technique. An augmented customer journey can be useful to service designers for mapping user experience, supporting the service design process. Moreover, EEG and eye tracking data quantify directly user experience in terms of factors, such as attention levels, eye blinks and pupil size. The advantage of this approach for in situ data collection lies in the ability to infer customers' implicit awareness throughout a customer journey.

The unique contribution is the use of quantitative, empirical and extensive measures within service design, converted into customer journey visualizations thus further extending existing service designers' tools. An ongoing study revealed that only 1 out of 164 analyzed SD tools and methods is quantitative.

The next step is the implementation of a platform for experience monitoring and customer journey visualization in application domain, as well as the exploration of various visualization of eye and EEG data for the given domain. The ultimate goal of our approach is to impact the service design community by validating potential influences of the augmented customer map in service designers.